\documentclass[aps,prl,float,twocolumn,a4paper,amsmath,amssymb]{revtex4}
\usepackage[dvips]{color}
\definecolor{g-blue}{rgb}{0.83,0.95,1}
\definecolor{g-yellow}{rgb}{1,1,0.7}
\definecolor{g-green}{rgb}{0.9,1,0.9}
\definecolor{green}{rgb}{0,0.6,0}
\definecolor{cyan}{rgb}{0,0.7,0.7}
\definecolor{black}{rgb}{0,0,0}
\definecolor{grey}{rgb}{0.4 ,0.4 ,0.4 }

 \usepackage{graphicx}
 \usepackage{amsmath,bm,epsfig}
\usepackage{verbatim}
\def \ed {\end{document}}
\def\Fbox#1{\vskip1ex\hbox to 8.5cm{\hfil\fboxsep0.3cm\fbox{%
  \parbox{8.0cm}{#1}}\hfil}\vskip1ex\noindent}  

\def\be{\begin{equation}}\def\ee{\end{equation}}
\def\bea{\begin{eqnarray}}\def\eea{\end{eqnarray}}
\def\bse{\begin{subequations}}\def\ese{\end{subequations}}
\newcommand{\BE}[1]{\begin{equation}\label{#1}}
\newcommand{\BEA}[1]{\begin{eqnarray}\label{#1}}
\newcommand{\BSE}[1]{\begin{subequations}\label{#1}}

\let \= \equiv \let\*\cdot \let\~\widetilde \let\^\widehat \let\-\overline

  \def\1{\bm1} 

\def\<{\left\langle}    \def\>{\right\rangle}
\def\({\left(}          \def\){\right)}
 \def \[ {\left [} \def \] {\right ]}

\newcommand{\B}[1]{{\bm{#1}}}
\newcommand{\C}[1]{{\mathcal{#1}}}    

\renewcommand{\sb}[1]{_{\text {#1}}}  
\renewcommand{\sp}[1]{^{\text {#1}}}  
\newcommand{\Sp}[1]{^{^{\text {#1}}}} 

\begin{document}

\title{Enhancement of intermittency in superfluid turbulence}
\author{Laurent Bou\'e, Victor L'vov, Anna Pomyalov, Itamar Procaccia}
\affiliation{Department of Chemical Physics, Weizmann Institute of Science, Rehovot 76100, Israel}

\begin{abstract}

We consider the intermittent behavior of superfluid turbulence in $^4$He. Due to the similarity in the nonlinear structure of the two-fluid model of superfluidity and the Euler and Navier-Stokes equations one expects the scaling exponents of the structure functions to be the same as in classical turbulence for temperatures close to the superfluid transition $T_\lambda$ and also  for $T\ll T_\lambda$. This is not the case when mutual friction becomes important.   Using shell model simulations, we propose that for an intermediate regime of temperatures, such that the density of normal and superfluid components are comparable to each other, there exists a range of scales in which the effective exponents indicate stronger intermittency.  We offer a bridge relation between these effective and the classical scaling exponents. Since this effect occurs at accessible temperatures and Reynolds numbers, we propose that experiments should be conducted to further assess the validity and implications of this prediction.
   \end{abstract}
\date{\today}
\maketitle

 \noindent
\textbf{Introduction}. Non-equilibrium systems are {often} characterized by relatively quiescent periods which are interrupted by rare but violent changes of the physical observables.  This phenomenon which is referred to as ``intermittency'' is very generic as it appears in many different stationary random processes. 

It is believed~\cite{Fri}, that highly excited   hydrodynamic turbulence possesses universal statistics in the  inertial interval of scales $L\gg r \gg \eta$. Here   ~$L$ is the of energy  input   scale, e.g. due to instabilities of high velocity flows,  and  $\eta$ is the viscous  energy dissipation scale.  The intermittent behavior of space homogeneous fully developed turbulence of incompressible fluids can be studied in terms of
  the  {\emph{velocity structure functions}}~\cite{Fri}
 \begin{eqnarray}\label{SF}
S\sb{p}( r) \equiv \langle |  \B  v(\bm R,t) - \bm v(\bm R + \bm r,t)| ^p \rangle \propto r^{\xi\sb{p}} \  .
\end{eqnarray}
Here~$\langle ... \rangle$ denotes a time averaging.  In the inertial interval, the functions $S\sb{p}( r)$ are scale invariant and they are characterized by   the~$p$-order scaling exponents~$\xi_p$.

 Accepting  Richardson's idea 
 of a step-by-step energy cascade in the inertial interval and assuming that the energy flux over the scales~$\varepsilon$ is the only relevant parameter,  Kolmogorov proposed from dimensional reasoning  that  $ \displaystyle
S_p\Sp{K41}( r) \simeq (\varepsilon r)^{p/3}$ \  i.e.  $\xi_p \equiv \xi_p\Sp{K41}=p/3$.
 In particular, this means that $S_p(r)/[S_2(r)]^{p/2}$ should be $r$-independent, like in Gaussian statistics. However this scenario is not borne out by experiments~\cite{Fri}: the observed values $\xi\sp{exp}$ for $p>3$  deviate down from $p/3$ and ther atios $S_p(r)/[S_2(r)]^{p/2}$ increase with $r$.  Therefore  the probability of observing large values of small scale velocity fluctuations (that mainly contribute to $S_p(r)$ with large $p$) significantly   exceeds its normal K41 level, i.e. the small scale turbulent flow is strongly intermittent. In spite of fifty years efforts intermittency still requires   deeper understanding. In particular there is  no  analytical framework to compute   $\xi\sb{p}(p)$.

 Turbulence in superfluid~$^3$He and~$^4$He  has emerged  as a hot topic of interest merging the traditional fluid mechanics  with the low-temperature physics.  Unlike classical fluids, where the circulation around the vortices is a   dynamical variable,   in superfluids it is quantized to integer values of $\kappa=2\pi \hbar /m$.  Here $\hbar$ is the Plank's constant and
 $m$ is the mass of a superfluid particle.  This leads to the appearance of   an additional  length scale  $\ell$,  the mean inter-vortex distance. For motions with scales $r\gg \ell$  superfluids can be described as two inter-penetrating entities with two different velocities: $\B v\sb{s}$ (inviscid superfluid component) and $\B v\sb{n}$ (normal component), and temperature dependent densities  $\rho\sb{s}(T)$ and $\rho\sb{n}(T)=\rho -\rho\sb s(T)$  interacting via  the so-called mutual friction force~\cite{LT}.

One of the most interesting questions concerns the similarities and differences between    superfluid  and classical turbulence. The pioneering experiments of Maurer and Tabeling suggest that intermittency would also be present in turbulent superfluid~$^4$He~\cite{tabeling}.   This crucial observation    implies that the existence of   intermittency  is not necessarily related to the exact structure of the Navier-Stokes equations and thus superfluid turbulence   offers an additional  way to investigate its physical mechanisms.  Unfortunately, intermittency in superfluids has received far less attention than in classical turbulence~\cite{leveque2012}.

 The aim of this Letter is to compare intermittency   in $^4$He-superfluid turbulence  with  turbulence in normal fluids. An important parameter characterizing superfluid turbulence is  the ratio ~$\rho \sb s /\rho \sb n$. For $T$ slightly below the phase transition temperature $T_\lambda$, when $\rho \sb s /\rho \sb n\ll 1$, one can neglect the  presence  of the superfluid component and the statistics of turbulent superfluid $^4$He  is expected to be close to that  of classical  fluids. We also expect similar inertial range behavior of  classical and   superfluid turbulence  for~$T\ll T_\lambda$, when $\rho\sb n \ll \rho \sb s$,  due to the inconsequential role played by  the normal component,  see, e.g.~\cite{leveque2012}. Moreover (and less trivially)
 the intermittent scaling exponents $\xi_p$ have to be the same in classical and low-temperature superfluid turbulence.   This is because the nonlinear structure of the equation for the superfluid
 component is the same as the Euler equation, and dissipative mechanisms are believed to be irrelevant. 

Based on the above reasoning, we focus   in on the    range of $T\simeq (0.9\div 0.8)T_\lambda$, where $\rho\sb s \sim \rho \sb n$  and   compare it to  the  high and low $T$  limits, when  $\rho\sb s/\rho =0.1$ and $0.9$,  respectively.  Our main tool is numerical simulations of superfluid turbulence in the two-fluid model~\cite{LT}. We utilize a shell-model approximation, which is known to provide accurate
scaling exponents in the classical case when the parameters are cleverly chosen~(see, e.g.~\cite{GOY,Sabra,S2}). In particular, shell models with the right parameters exhibit anomalous scaling exponents $\xi_p\sp{num}$ that agree well with experimental values $\xi_p\sp{exp}$: $\xi_2\sp{num}=0.72$ vs. $\xi_2\sp{exp}=0.70$; $\xi_4\sp{num}=1.26$ vs. $\xi_4\sp{exp}=1.28$, etc.

The main result  of the Letter is as follows:  For temperatures such that~$\rho\sb s / \rho\lesssim 0.1$ and~$\rho\sb s / \rho \gtrsim 0.9$, the scaling behavior of superfluid turbulence is practically the same as in classical fluids. However,  for temperatures  such that~$\rho\sb s / \rho\simeq 0.5\div 0.75$, we discovered  the emergence of a   wide (up to 3  decades) interval of scales  characterized by  substantially different  effective scaling exponents~$\~\xi_p\ne \xi_p\sp{num}$.   For example,  when $\rho_s=\rho_n$  the  value of $\~\xi_2$ diminishes from its intermittent value $\xi_2\sp {num}\approx 0.72$  down to $\tilde \xi_2\approx 0.67$, close to the K41 value $\xi_2\Sp{K41}=2/3$. At the same time, the  exponents  $\~\xi_p$ for $p>3$ deviate further from the K41 values: $\~\xi_p<\xi_p\sp{num}< \xi_p\Sp{K41}$.  This fact can be considered as \emph{enhancement of intermittency in superfluid turbulence} compared to the    classical case.

In addition, based on   numerical observations that the probability distribution function (PDF) of the normalized shell energy  is temperature  independent, we suggest the  relation between effective  and classical exponents:
 \begin{equation}\label{res1} \tilde \xi_p = \xi_p\sp {num} - \frac p2 (\xi_2\sp{num}-\tilde \xi_2)\approx \xi_p\sp{num} - 0.026\,  p\,, \end{equation}
  which  is in good agreement with  observed  values of $\tilde \xi_p$.

\begin{figure*}[t]
\includegraphics[width=0.315\linewidth]{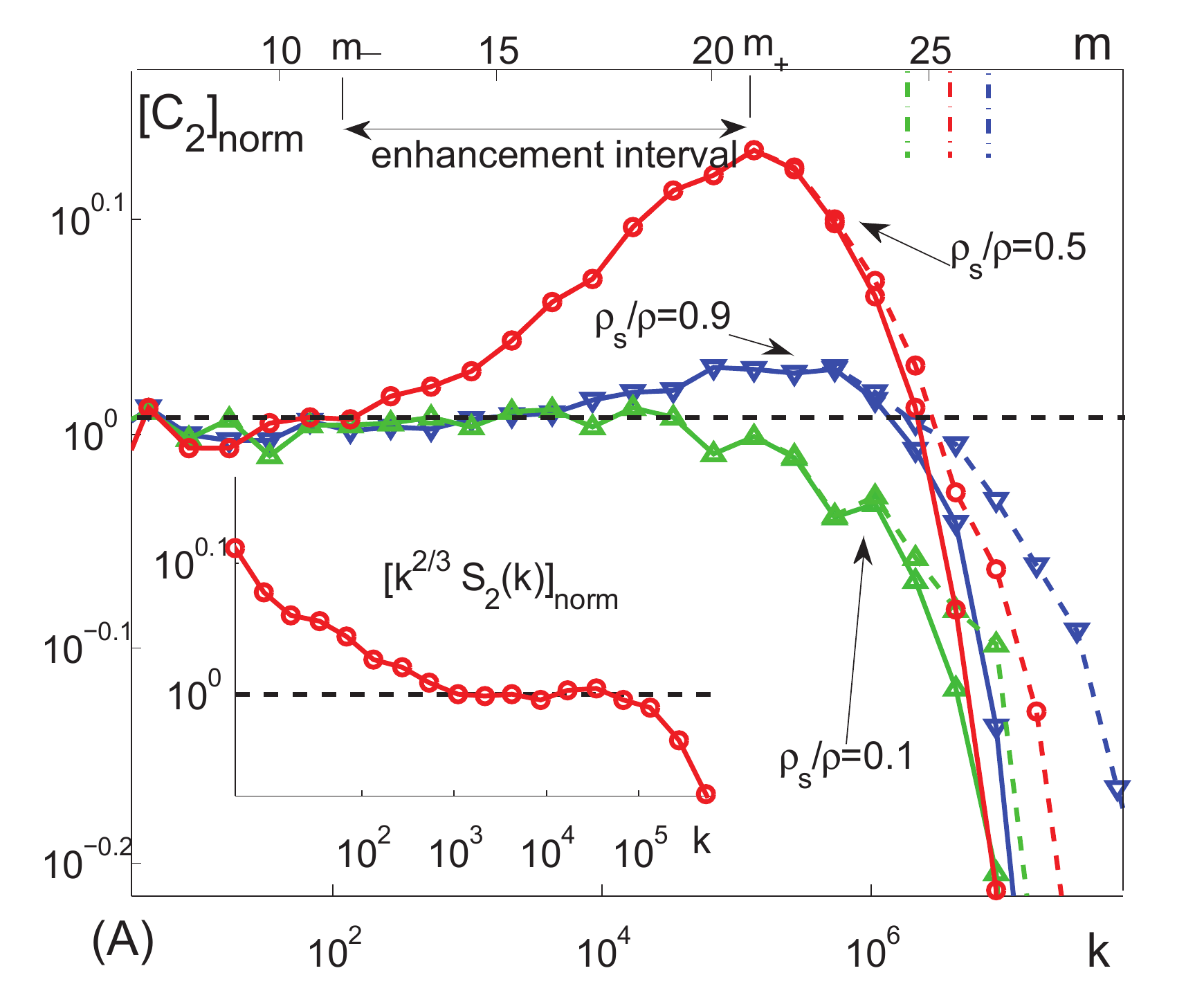}
\includegraphics[width=0.33\linewidth]{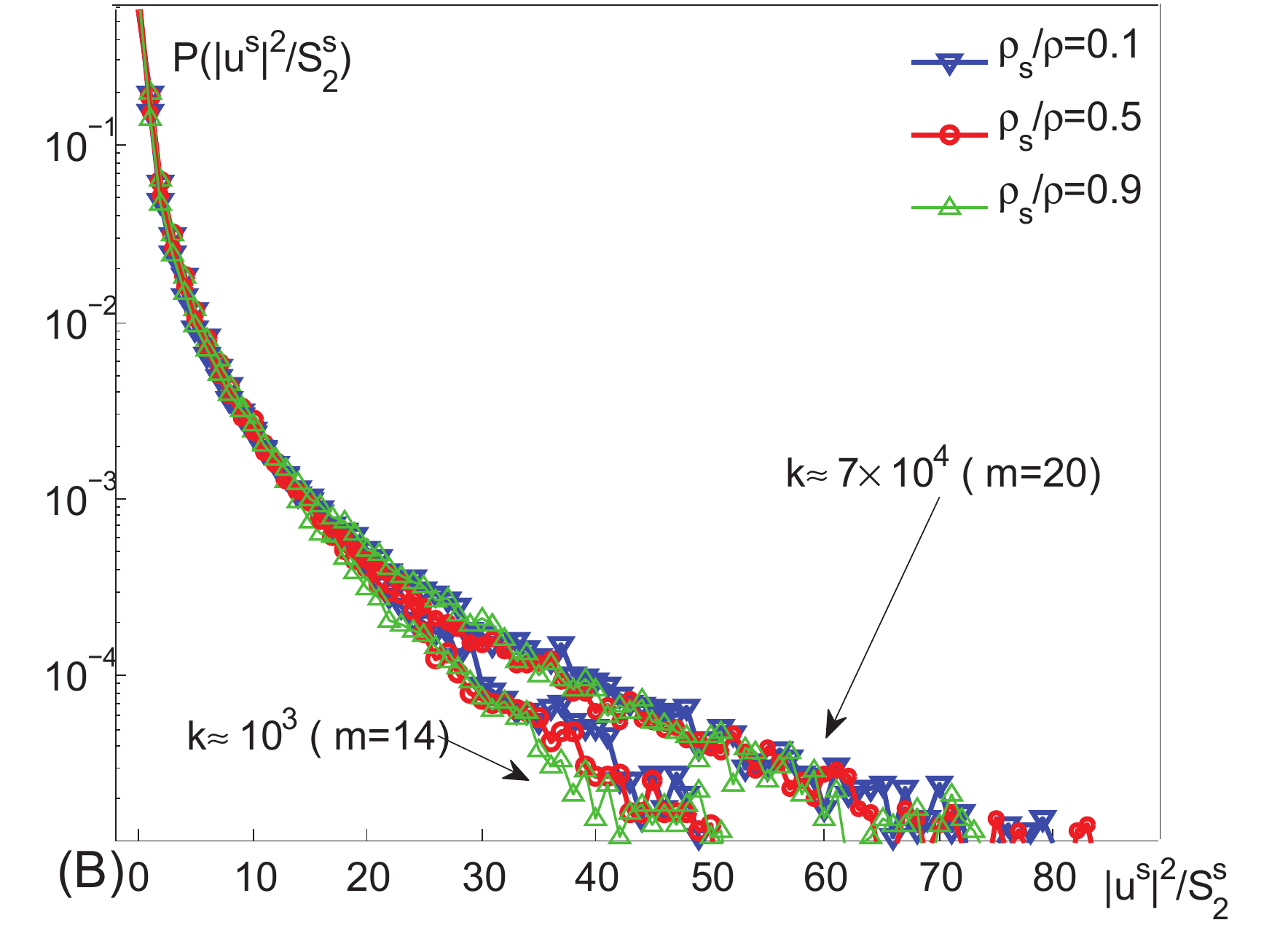}
\includegraphics[width=0.335\linewidth]{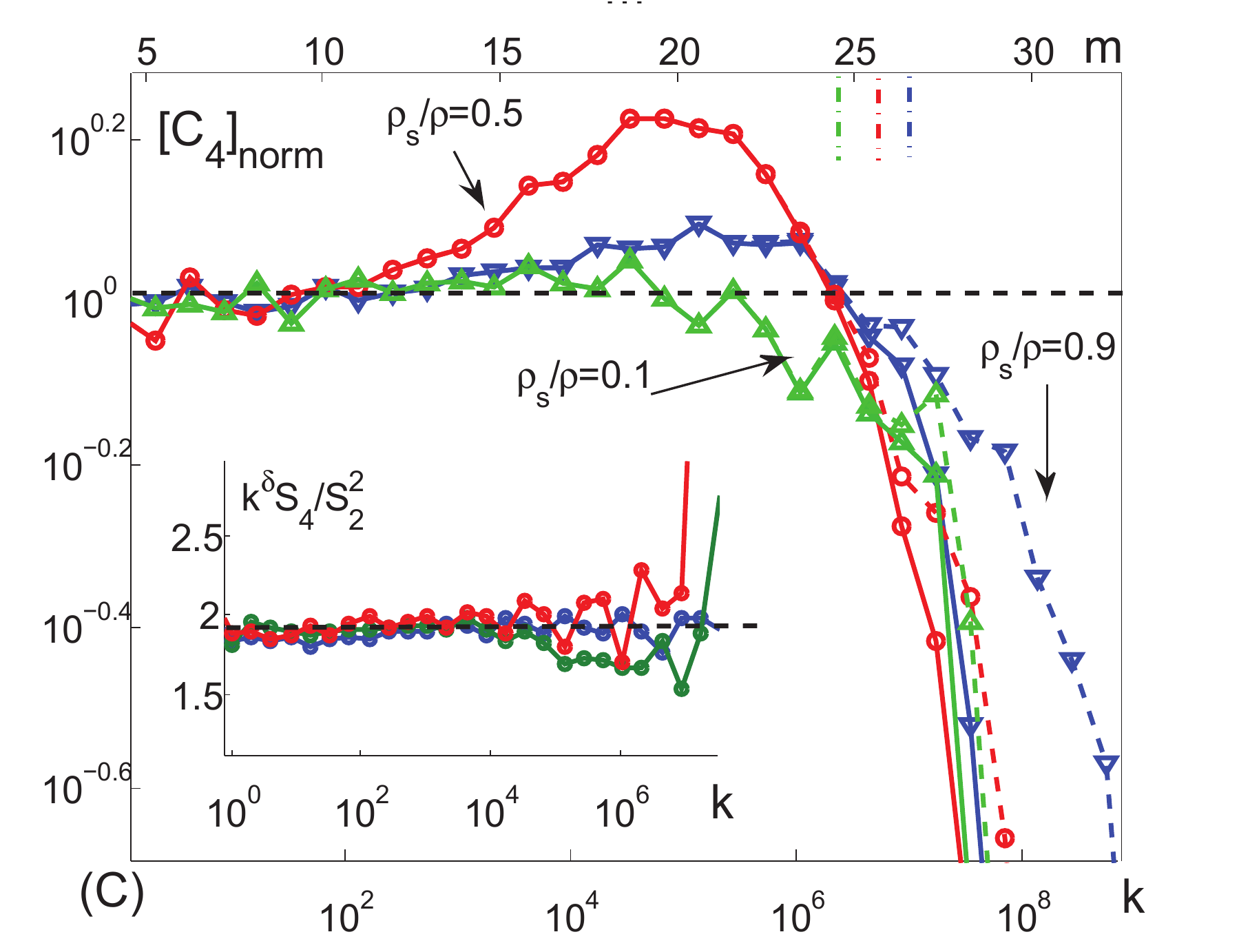} \vskip -0.3 cm
\caption{\label{f:1} Panel A: The compensated structure functions $C_2\sp{n,s}(k_m)$   for  $\rho\sb s/\rho=0.1, 0.5$ and  $0.9$, normalized by its mean value over shells with $m=7-14$.   Inset: $k^{2/3}\,S_2(k_m)$ normalized by its mean value over $m=15-20$ for $\rho\sb s/\rho=0.5$. Solid lines -- normal component, dashed lines -- superfluid components. The upper axis is marked with the shell numbers, the lower axis shows the corresponding wavenumbers.  Panel B: Demonstration of the temperature-independence of the PDF's for shells $m=14$ and $m=20$. Panel C: The normalized compensated  $C_4\sp{n,s}(k_m)$.  Inset: The collapse of $ k^\delta S_4(k)/[S_2 (k)]^2$, compensated using classical   exponents $\delta\equiv 2 \xi_2\sp{num}-\xi_4\sp{num}\approx 0.184$. The horizontal black dashed lines serve to guide the eye only. Color code and line labeling is the same in all panels. Vertical dot-dashed lines in Panels A and C  denote the intervortex scale $1/\ell$ (see~\cite{WB}).  }
 \end{figure*}
\vskip 0.3 cm
\noindent
\textbf{Numerical procedure}. Shell models of superfluid turbulence~\cite{WB,BLPP}
are simplified caricatures
of the Navier-Stokes and Euler equations   in $\B k$-representation:
 \begin{subequations}\label{SM}
 \begin{eqnarray}\label{SMn}
   \Big[\frac{d}{dt}+\nu\sb n  k_m^2\Big]u\sp {n} _m& =&\mbox{NL}[u\sp {n} _m]+   F_m\sp {n} +f_m\sp {n} \, ,\\
\label{SMs}
 \Big[\frac{d}{dt}+\nu\sb s  k_m^2\Big]u\sp {s} _m  & =&\mbox{NL}[u\sp {s}_m]-F_m\sp {s} +f_m\sp {s}\, ,\\ \label{NL}
  \mbox{NL}[u_m ]&=& i \Big(a  k_{m+1} u_{m+2}u^*_{m+1}  \\ \nonumber
&& \hskip - 2,5 cm +b k_m  u_{m+1} u^*_{m-1}-c k_{m-1}u_{m-1}u_{m-2}\Big)\,,\\ \label{F}
 && \hskip - 4 cm F_m\sp{n,s}=   \alpha\sb{n,s}\Omega\sb s(u\sp {s}_m-u\sp {n} _m)\,, \quad \Omega\sb s^2\equiv \sum_m k_m^2 |u_m\sp {s}|^2   \ .
 \end{eqnarray}
\end{subequations}
They mimic the statistical behavior of $\B k$-Fourier
components of the turbulent superfluid and normal velocity fields   in the entire shell of wave vectors $k_m < k < k_{m+1}$ by   complex shell
velocity $u_m\sp{n,s}$. The shell wave numbers  are chosen as a geometric
progression $k_m = k_0 \lambda^m$, where~$m=1,2,\dots\,M $ are the shell indexes, and we have used  the shell-spacing parameter $\lambda=2$  and~$k_0=1/16$.
In order to resolve a large range of scales we used  a total of ~$M=36$  shells. This allows us to clearly  resolve   subregions with different scaling behavior.

Similarly to the Navier Stokes equation, the  NL$[u_m]$   term in Eq.~\eqref{NL}   is quadratic in velocities, proportional to $k$ and conserves (in the force-less, inviscid limit) the kinetic energy 
$E = \frac 12 \sum_m   |u_m |^2$,
provided that ~$a+b+c=0$. We used the Sabra  version~\cite{Sabra} of   NL$[u_m]$ with the traditional (and physically motivated choice)~$b=c=-a/2$, which ensures~\cite{GOY,Sabra} that   $\xi_p\sp {num}$ are close to   $ \xi_p\sp {exp}$.

In Eq.~\eqref{SMn}, $\nu\sb{n}$ is the kinematic viscosity of normal component.
It is known that vortex reconnections   act as an energy sink in superfluids,  which we   accounted for (admittedly in a rough manner) by    adding in Eq.~\eqref{SMs} an effective superfluid viscosity $\nu\sb s \ll \nu\sb{n}$. The details of this complicated and otherwise still not fully understood  dissipation process go beyond the scope of this Letter. In simulations we used  $\nu\sb s=10^{-12}$ and $\nu\sb n=10^{-10}$.

The coupling terms~$F_m\sp{n,s}$ \cite{WB,BLPP},  given by  Eq~\eqref{F}, account
 for the mutual friction between normal and superfluid velocities.   This dissipative force  is proportional to    parameters~$\alpha \sb s(T)= \alpha$ and~$\alpha \sb n(T) = \alpha \rho\sb s /\rho\sb n$.

The Gaussian $\delta$-correlated in time  forces $f\sp{n,s}_m$ were applied   only  for $m=1,2$.  Eqs~\eqref{SM} were solved using 4-th order Runge-Kutta method.
 The chosen values of $\rho\sb s/\rho$ and corresponding values of $T/T_\lambda$ and $\alpha$ (interpolated using the experimental data in~\cite{ExpData}) are shown below:
 \begin{table}[h]
\begin{tabular}{| c||c|c|c|c|c|c|c|c|c|c|}
\hline
$\rho\sb{s}/{\rho}$ & 0.1 & 0.25  & 0.4   & 0.5   &   0.6   & 0.65  & 0.7  & 0.75  &  0.9  \\
\hline
$T/{T_{\lambda}}$ & 0.99  & 0.97 & 0.93  & 0.90  &   0.87 & 0.84  & 0.82 & 0.79  &  0.677 \\
\hline
$\alpha$          & 1.09 & 0.51 & 0.314 & 0.25  & 0.22  &   0.175 & 0.155 & 0.14 &  0.07  \\
\hline
$\alpha \rho/\rho\sb n$ & 1.21 & 0.68 & 0.52 & 0.50 &  0.49 & 0.50 & 0.52 & 0.54 &  0.67 \\
 \hline
\end{tabular}
 \end{table}

\noindent
\textbf{Results and discussions}. First of all, we study the scaling behavior of the velocity structure functions~\eqref{SF}, which in the shell models can be written as 
\begin{equation}   S_{p}\sp{s,n} (k_m) \equiv \langle |u_m\sp{s,n}|^p\rangle\,, \quad  C_p(k_m)\equiv k_m^{\xi_p\sp{num}}S_p(k_m)\ . 
 \end{equation}
 We defined for convenience the structure functions $ C_p(k_m)$  compensated by the classical anomalous scaling.

Figure~\ref{f:1}A shows $C_2(k_m)$.  The almost horizontal lines for $\rho\sb s/\rho =0.9$ and $0.1$, indicate the usual intermittent scaling behavior. However for  $\rho\sb s/\rho=0.5$  one detects the emergence of a wide  \emph{enhancement interval} with the energy accumulations and where the scaling exponent is close to~$\~ \xi_2\simeq 2/3$~\cite{footnote}.  This is clearly seen in the inset, where $S_2(k_m)$ is compensated with K41 scaling.  At first glance, the  scaling    with  $\~ \xi_2\simeq \xi_2\Sp{K41}=2/3$ can be considered as the normal K41 scaling in superfluids. In this case one would expect that the PDF  of normalized energy in $m$-shell  $P\sp {n,s}\big[ |u_m\sp {n,s}|^2 / S_2 \sp{n,s}(k_m)\big ]$ should collapse for all $m$. This is, however, NOT the case here.  Figure~\ref{f:1}B  shows that the PDFs for different $m$ are in fact very different from each other.  The much longer tails associated with larger values of $k_m$ reveal strong intermittency effects.
This is clearly seen in Fig.~\ref{f:1}C  where   $C_4\sp{n,s}(k_m)$, is almost flat for $\rho\sb s/\rho =0.9$ and $0.1$, but show a  different scaling for $\rho\sb s/\rho=0.5$ with an effective exponent $\~\xi_4\approx 1.21 < \xi_4\sp{num}=1.256 < \xi\Sp{K41}= 4/3$.  The next order structure functions, $S_6(k)$,  $S_8(k)$,.. demonstrate the same type of behavior where the effective  exponents $\~\xi_p$ deviate  even  further  away  from the K41 values  than  the intermittency exponents $\xi_p\sp{num}$ in classical fluids.

Another important observation in Fig.~\ref{f:1}B  is that the  PDFs are practically temperature independent. Then   knowing $\xi\sp{num}_p$ and $\~\xi_2$, we can predict with very good accuracy all the other effective exponents $\~\xi_p$. Indeed,  comparing the following expressions for $S_p\sp{n,s}(k_m)$ \\  
 $\displaystyle  ~~~~~S_p\sp{n,s}(k_m)= \int _0 ^\infty  \C P\sp {n,s}\big[ |u_m\sp {n,s}|^2 / S_2 \sp{n,s}(k_m)\big ] |u\sp{n,s}_m|^p d u\sp{n,s}_m$\\
in the regions  of different scaling, one arrives after some simple algebra to  Eq.~\eqref{res1}.  An immediate consequence of Eq.~\eqref{res1} is that   ratio    $S_p(k)/[S_2(k)]^{p/2}$ is the same in the regions with $\xi_p \sp{num}$ and $\~\xi_p$ scaling. This prediction fully agrees with observations (see e.g. inset in  Fig.~\ref{f:1}C  for $p=4$)  and thus  lends  strong support  to  Eq.~\eqref{res1}.

\begin{figure}[b]\vskip -0.45 cm
\includegraphics[width=0.62\linewidth]{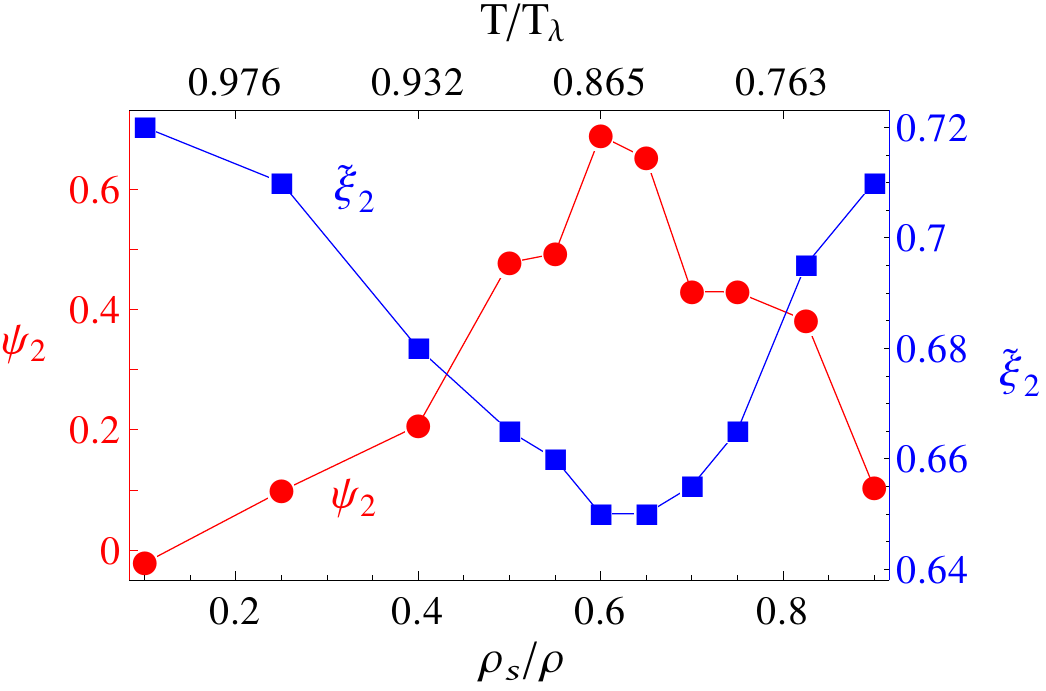}\vskip -0.4 cm
\caption{\label{f:2}  $T/T_{\lambda}$- and $\rho_s(T)/\rho$-dependencies (upper and lower axes) of the enhancement magnitude $\Psi_2$, left axis,  and the exponent $\~ \xi_2$, right axis.  The maximum of $\Psi_2\approx 0.7$ and minimum of $\~\xi_2\approx 0.65$ are reached at $T\approx 0.865 T_\lambda$ ($\rho_s\approx 0.6\rho$).   }
  \end{figure}

To  quantify  the intermittency enhancement we display in Fig.~\ref{f:2} the  $T$-  and $\rho\sb s/\rho$-dependence of $\~\xi_2$ together with the dimensionless magnitude $ \Psi_2$ of this effect, defined as
$\displaystyle   \Psi_2\equiv \log_2\{ [\max_m C_2(k_m)/  \langle C_2(k_m)\rangle ]\}$.
Here   $ \langle C_2(k_m)\rangle $ is the mean value of  $C_2(k_m)$ on the plateau.
Denote by $m_\pm$ the left and right edges of the  enhancement interval  and its  extent as $k_{m_+}/k_{m_-}= 2^{\Delta m}, \ \Delta m \equiv (m_+-m_-)$. Then,  using data in Fig.~\ref{f:2} and     computing the ratios   $\Psi_2(T)/ [\xi_2\sp{num}-\~\xi_2(T)] \equiv \Delta m$ for all   $T$,   we found that
$\Delta m \simeq 8\div 10$   practically independent of $T$.
In other words the extent of the  enhancement interval   $k_{m_+}/k_{m_-}=2^{\Delta m}\simeq 250\div 1000$.

As expected, significant  intermittency enhancement takes place for $\rho\sb s\sim  \rho\sb n$  [or more precisely    $\rho\sb s \simeq (0.4\div 0.8)\rho$ -- see Fig.~\ref{f:2}] when the two-fluid model~\eqref{SM} differs mostly from a single classical fluid. Normal and superfluid   components in Eqs.~\eqref{SM} are coupled by the mutual friction force $F_m\propto \alpha (u_m\sp s - u_m\sp n)$ which causes the energy exchange between normal and superfluid components and the energy dissipation. Thus these two phenomena are the only possible reasons for the observed  intermittency enhancement and both  disappear in case of full velocity locking $u_m\sp s(t)  = u_m\sp n(t)$. Velocity locking  can be statistically quantified by the cross-correlation coefficient of normal and superfluid velocities:
$\displaystyle \C K_m\equiv \langle [u_m\sp {s*}u_m\sp{n} + u_m\sp {s }u_m\sp{n*}]\rangle/\langle [u_m\sp {s*}u_m\sp{s} + u_m\sp {n }u_m\sp{n*}]\rangle$.
For  $\C K(k_m)=0$,  the superfluid and normal components are statistically  independent subsystems, while for  $\C K(k_m)=1$
they are completely locked  to each other. Clearly only for $0<\C K(k_m)<1$    we can expect intermittency enhancement.  A result of~\cite{LNS} for the correlation coefficients can be reformulated for shell models as:
 $   \displaystyle
 \C K_m \simeq\frac{2[\alpha\sp{n} S_{2,m}\sp{s}   + \alpha\sp{s} S_{2,m}\sp {n}  ] \big / [ S_{2,m}\sp{s}  + \!S_{2,m}\sp{n}  ]  }{\alpha\sb{s}+\alpha\sb{s}+ k_m \big[\nu\sb n   k_m +  S_{2,m}\sp{s}  + \!S_{2,m}\sp{n}    \big]/\sqrt{\langle \Omega\sb{s}^2\rangle }}\ .  $ \\
   \noindent
 Here $S_{2,m}\sp{n}\equiv S_{2}\sp{n}(k_m)$.  Analysis of this equation  shows that the $v\sp s_m$ and $v\sp n_m$ are   locked when $k_m \eta \ll 1$ and become uncorrelated when $ k_m\simeq  \big ( \alpha \rho/ \rho\sb n\big )^{3/2}/ \eta $. As one sees in the Table, for all $T$ in our simulation   $ \alpha \rho/ \rho\sb n \simeq 0.5 \div 1.2$. Therefore one should expect   velocity locking      for all  $T$ and for all  $k_m\eta\lesssim 1$ including the enhancement range. This prediction agrees with our results.

This means that  the   origin for the   variations of the scaling \emph{is not in the enhancement range} but rather  \emph{in the vicinity of the viscous cutoff}, where  $k_+ \sim k_\eta$.  This conclusion is strongly supported by our simulations with different  $T$, $M=28 \div46$    and $\nu\sb n/\nu\sb s \simeq  10^2\div 10^6$, in all of which the extent of the enhancement interval  remains unchanged, $\Delta m\simeq 8\div 10$,  and always $k_+ \sim k_\eta$.

One concludes that a key role should be played by the energy dissipation due to the   mutual friction  $D_m\Sp{MF}= \alpha \rho\sb s  \Omega\sb s |u\sp s_m - u\sp n _m|^2 $, as follows from
 Eqs.~\eqref{SM}. Clearly,  $D_m\Sp{MF}\to 0$ for $T\to T_\lambda$ (because $\rho\sb s \to 0$) and for
 $T\to 0$ (because $\alpha \to 0$) and   has a   maximum at $T\simeq
0.80\div 0.85$, i.e. in the range, where the intermittency enhancement is observed.
Moreover, in this range, the total mutual friction dissipation $\sum_m D_m\Sp{MF}$ is close to
(or even larger  than) the total viscous dissipation $\nu \sb n \sum_m k_m^2 S_{2,m}\sp n$.
Importantly, $D_m\Sp{MF}$ has a maximum at $k_m $, close to that of the viscous dissipation. As a result, for $T\simeq
(0.80\div 0.85) T_\lambda $, the mutual friction  between superfluid and normal components  markedly   drives  the system  away  from the simple one-fluid  behavior primarily  around $k\sim k_\eta$, leading to a sharper decay of the energy spectrum $S_{2,m}\sp{n,s}$. In turn, this may  give rise to  a bottleneck  effect leading to an  energy accumulation near $k_\eta$. This phenomenon shares similarities with the bottleneck in classical fluids~\cite{BN1}
but differs in important aspects: e.g. as  in classical fluids the width of the energy accumulation range is almost independent of underlying mechanism and only its magnitude varies. In our case, however, the extent of the enhancement interval is much larger, about 3 decades, compared to one decade in classical fluids. A possible  reason for such an   extended   enhancement interval is  the energy transfer from superfluid to normal component, caused by mutual friction,  and it is also  notable in the same $T$ range.

  A comprehensive study  into the quantitative aspects of  this  enhancement of intermittency  is beyond the scope of the Letter.  It will require an extensive analysis of the time dependencies of $ u\sp s_m(t)$ and  $u\sp n _m(t)$ in order to clarify the  role of the  mutual friction on the statistics of outliers;  the main contributors to the fat tails of the PDFs which are responsible for the intermittency~\cite{out}.

 \noindent
\textbf{Summary}. Our shell-model simulations of the large-scale superfluid  turbulence in  $^4$He  indicate that for high- and low-$T$-limits, when $\rho\sb s<0.1\rho$ and $\rho\sb s> 0.9 \rho$ anomalous scaling exponents of superfluid turbulence are practically the same as their classical counterparts, as follows from our theoretical reasoning   and agrees with numerical and laboratory observations~\cite{leveque2012}.  The shell-model simulations have allowed us to  uncover  considerable   $T$-dependent variations of the apparent scaling over the \emph{enhancement  interval}  of scales about  3  decades  when $\rho\sb s \sim \rho \sb n$.   This   scaling    can be characterized by the  effective scaling exponents $\~\xi_p(T)$ that deviate further from the normal K41 values as compared to the exponents of classical fluids.  Furthermore if  the entire inertial interval is shorter than  the enhancement interval,   only the effective scaling    is observed.
We also suggested   Eq.~\eqref{res1}   relating $\~\xi_p(T)$ to $\xi_p$.

 While our  interpretation  indicates  the existence of  a   form of the energy bottleneck in the two-fluid system, the phenomenon definitely calls for further investigation. DNS of superfluid turbulence oriented towards the study of the predicted $T$-dependent scaling over an interval  of about three decades  is nowadays realistic and  appears as particularly timely.  Even more importantly, the temperature range $T\simeq (1.7\div 1.9)\,$K,  where we forecast a change in the scaling behavior is accessible to laboratory experiments. We hope that the  results presented in this Letter will inspire new numerical and laboratory studies of intermittency in superfluid turbulence.

\acknowledgments
This work is supported by the EU FP7 Microkelvin program  (Project No.~228464) and by  the U.S. Israel Binational Science Foundation.


\begin{thebibliography}{99}
\bibitem{footnote} Notice that we observe the energy accumulation for scales, larger then intervortex distance $\ell$, while in~\cite{BB} it happens at  smaller scales.
\end{thebibliography}
\end{document}